\documentclass[11pt,twoside]{article}
\usepackage{asp2010}
\usepackage{aas_macros}
\resetcounters

\begin{document}

\title{Efficient Parallelization for AMR MHD Multiphysics Calculations; Implementation in AstroBEAR}
\author{Jonathan~Carroll-Nellenback$^1$, Brandon~Shroyer$^1$, Adam~Frank$^1$, and Chen~Ding$^2$}
\affil{$^1$Department of Physics and Astronomy, University of Rochester, Rochester, NY 14620}
\affil{$^2$Department of Computer Science, University of Rochester, Rochester, NY 14620}

\begin{abstract}
Current AMR simulations require algorithms that are highly parallelized and manage memory efficiently.  As compute engines grow larger, AMR simulations will require algorithms that achieve new levels of efficient parallelization and memory management.  We have attempted to employ new techniques to achieve both of these goals.  Patch or grid based AMR often employs ghost cells to decouple the hyperbolic advances of each grid on a given refinement level.  This decoupling allows each grid to be advanced independently.  In AstroBEAR we utilize this independence by threading the grid advances on each level with preference going to the finer level grids.  This allows for global load balancing instead of level by level load balancing and allows for greater parallelization across both physical space and AMR level.  Threading of level advances can also improve performance by interleaving communication with computation, especially in deep simulations with many levels of refinement.  To improve memory management we have employed a distributed tree algorithm that requires processors to only store and communicate local sections of the AMR tree structure with neighboring processors.  
\end{abstract}

\section{Introduction}
The development of Adaptive Mesh Refinement \citep{Berger1984484, Berger198964} methods were meant to provide high resolution simulations for much lower computational cost than fixed grid methods would allow.  The use of highly parallel systems and the algorithms that go with them were also meant to allow higher resolution simulations to be run faster (relative to wall clock time).  The parallelization of AMR alogrithms, which should combine the cost/time savings of both methods is not straight forward however and there have been many different approaches \citep{Paramesh2000, Nirvana2008, Enzo2004, FTT}, .  While parallelization of a uniform mesh demands little communication between processors, AMR methods can demand considerable communication to maintain data consistency across the unstructured mesh as well as shuffling new grids from one processor to another to balance work load.  

In this paper we report the development and implementation of new algorithms for the efficient parallelization of AMR designed to scale to very large simulations. The new alogorithms are part of the AstroBEAR package for simulation of astrophysical fluid multi-physics problems \citep{Cunningham2009ApJS}.  The new algorithmic structrure described in this paper constitudes the development of version 2.0 of the AstroBEAR code. 

AstroBEAR like many other grid based AMR codes utilizes a nested tree structure to organize each individual refinement region.  However, as we will describe, unlike many other AMR codes, AstroBEAR 2.0 uses a distributed tree in which no processor has access to the entire tree but rather each processor is only aware of the AMR structure it needs to manage in order to carry out its computations and perform the necessary communications.  While currently, this additional memory is small compared to the resources typically available to a CPU, future clusters will likely have much less memory per processor similar to what is already seen in GPU's.  Additionally each processor only sends and receives the portions of the tree necessary to carry out its communication.

AstroBEAR 2.0 also uses extended ghost zones to decouple advances on various levels of refinement.  As we show below this allows for each level's advance to be computed independently on separate threads.  Such inter-level threading allows for total load balancing across all refinement levels instead of balancing each level independently.   Independent load balancing becomes especially important for deep simulations (simulations with low filling fractions but many levels of AMR) as opposed to shallow simulations (high filling fractions and only a few levels of AMR).  Processors with coarse grids can advance their grids simultaneously while processors with finer grids advance theirs.  Without such a capability, each level would need to have enough cells to be able to be distributed across all of the processors.  Variations in the filling fractions from level to level can make the number of cells on each level very different.  If there are enough cells on the level with the fewest to be adequately distributed, there will likely be far too many cells on the highest level to allow the computation to be completed in a reasonable wall time.  This often restricts the number of levels of AMR that can be practically used.  With inter-level threading this restriction is lifted.  Inter-level threading also allows processors to remain busy while waiting for messages from other processors.

In what follows we provide descriptions of the new code and its structure as well as providing tests which demonstrate its effective scaling.  In \ref{amr_alg} we review patch based AMR.  In section \ref{distributedtree} we will discuss the distributed tree algorithm, in section \ref{threading} we will discuss the inter-level threading of the advance, in section \ref{loadbalancing} we will discuss the load balancing algorithm, and in section \ref{performanceresults} we will present our scaling results.

\section{AMR Algorithm}\label{amr_alg}

\begin{figure}
 \caption{Example AMR mesh showing nested and adjacent grids as well as corresponding AMR tree showing parent-child and neighbor relationships.}
 \centering
  \includegraphics[width=.9\textwidth]{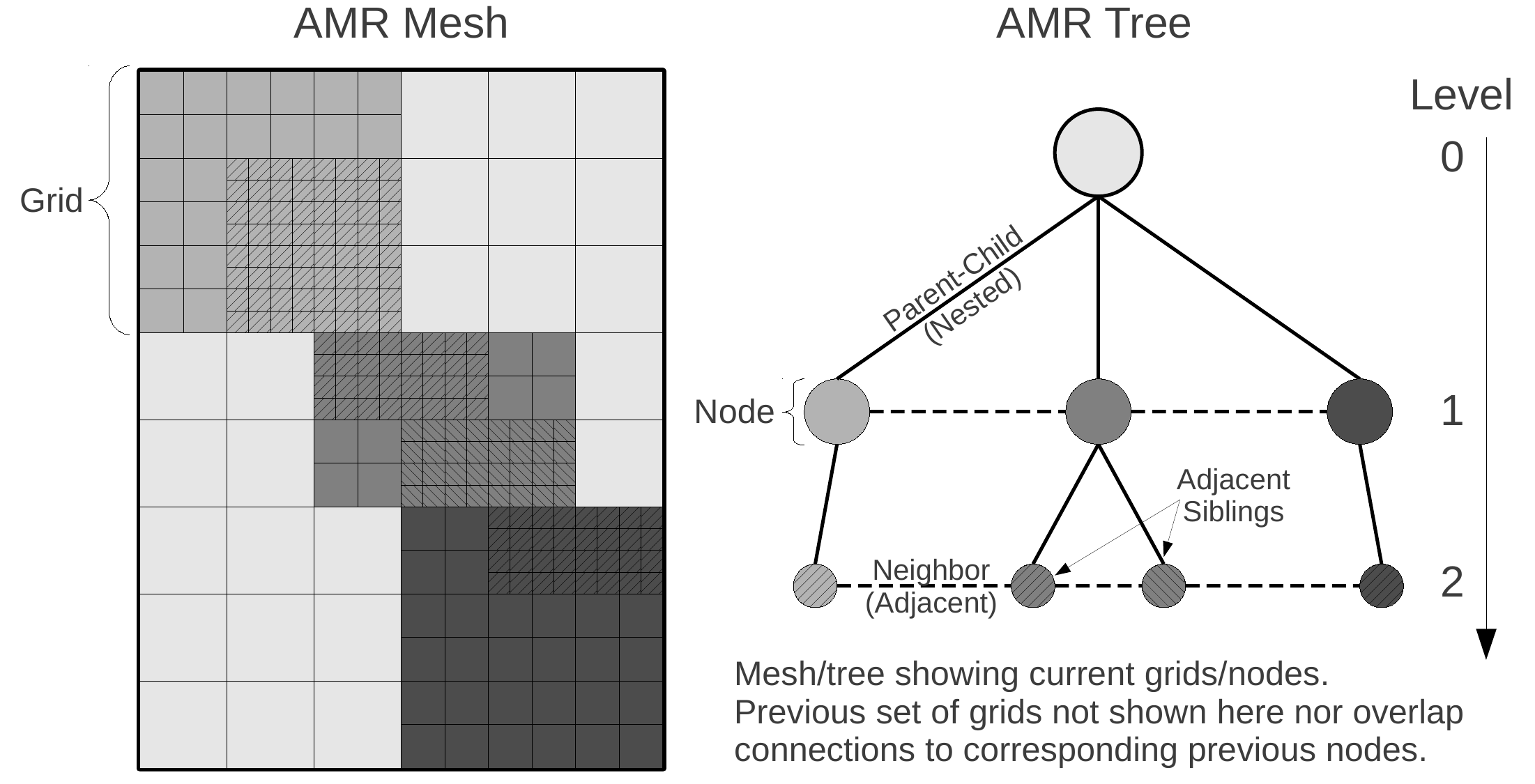} 
\label{meshtree}
\end{figure}

  Here we give a brief overview of patch based AMR introducing our terminology along the way.  The fundamental unit of the AMR algorithm is a patch or grid.  Each grid contains a regular array of cells in which the fluid variables are stored.  Grids with a common resolution or cell width $\Delta x_l$ belong to the same level $l$ and on all but the coarsest level are always nested within a coarser "parent" grid of level $l-1$ and resolution $\Delta x_{l-1} = R \times \Delta x_l $ where $R$ is the refinement ratio.  The collection of grids comprises the AMR mesh, an example of which is shown in figure \ref{meshtree}.  In addition to the computations required to advance the fluid variables, each grid needs to exchange data with its parent grid (on level $l-1$) as well as any child grids (on level $l+1$).  Grids also need to exchange data with physically adjacent neighboring grids (on level $l$).  In order to exchange data, the physical connections between parents, children, and neighboring grids are stored in the AMR tree as relational connections between nodes.  Each grid on each level of the entire AMR mesh has a corresponding node that is part of the AMR tree.  Thus there is a one to one correspondence between nodes and grids.  The grids hold the actual fluid dynamical data while the nodes hold the information about each grid's position and its connections to parents, children and neighbors.  Figure \ref{meshtree} shows one example of an AMR mesh made of grids and the corresponding AMR tree made of nodes.  Note that what matters in terms of connections between nodes is the physical proximity of their respective grids.  While siblings share a common parent, they will not necessarily be neighbors, and neighbors are not always siblings but may be 1st cousins, 2nd cousins, etc...  

  Additionally since the mesh is adaptive there will be successive iterations of grids on each level as the simulation progresses.  Thus the fluid variables need to be transferred from the previous iteration of grids to the current iteration.  Thus nodes can have ``neighbors'' within a 4 dimensional spacetime.  Nodes that are temporally adjacent (belonging to either the previous or next iteration) and spatially coincident are classified as preceding or succeeding overlaps respectively instead of temporal neighbors, reserving the term neighbor to refer to nodes of the same iteration that are spatially adjacent and temporally coincident.  Nodes on level $l$ therefore have a parent connection to a node on level $l-1$, child connections to nodes on level $l+1$, neighbor connections to nodes on level $l$ of the same iteration, and overlap connections to nodes on level $l$ of the previous or successive iteration in time.


\section{Distributed Tree Algorithm}\label{distributedtree}

Many current AMR codes store the entire AMR tree on each processor.  This, however, can become a problem for simulations run on many processors.  Let us first assume that each AMR grid requires $m$ bytes per node to store its meta data (ie 6 bytes to store its physical bounds for a 3D simulation and 1 byte to store the processor containing the grid).  We also assume that each grid requires on average $d$ bytes for the actual data.  If there are, on average, $n$ grids on each of $p$ processors, then the memory per processor would be $nd + nmp$.  The second term $nmp$ represents the metadata for the {\it entire} AMR tree.  

The memory requirement for just the nodes in the AMR tree without storing any connections becomes comparable to the local actual data when $p = d/m$.  If we assume a 3D isothermal hydro run where each cell contains $\rho, p_x, p_y, \& p_z$ with a typical average grid size of 8x8x8 then $p=\frac{8\times8\times8\times4}{(6+1)} \approx 293$.  While this additional memory requirement is negligible for problems run using typical cpus on 100's of processors, it can be become considerable when the number of processors $n > 10^3$.  Since it is expected that both efficient memory use and management will be required for ever larger HPC (high performance computing) clusters down the road, AstroBEAR 2.0 is designed to use a {\it distributed} tree algorithm. In this scheme each processor is only aware of the section of the tree containing nodes that connect to its own grids' nodes.  Additionally, new nodes are communicated to other processors on a proscriptive ``might need to know basis''.  Since maintaining these local trees as the mesh adapts is not trivial, we describe the process below.

\subsection{Maintaining the AMR Tree}

 Because of the nested nature of grids, neighbor and overlap relationships between nodes can always be inherited from parent relationships.  For example consider the neighbors of the $n^{th}$ iteration of a node's children.  The nested nature of the grids restricts each of the child's neighbors to either be a sibling of that child (having the same parent node and be of the same iteration), or to be a member of a neighbor's $n^{th}$ iteration of children.  Thus the neighbors of a level $l$ node's children (on level $l+1$) will always be a child of that level $l$ node's neighbors.  

  For parallel applications, the grids are distributed across the processors.  In addition to data for the local grids, each processor needs to know where to send and receive data for the parents, neighbors, overlaps, and children of those local grids.  This is information contained within the nodes.  In order for a processor to know where to send data, each one must maintain a local sub-tree containing its own "local" nodes ( corresponding to local grids) as well as all remote nodes (living on other processors) directly connected to the local nodes.  It is also possible, though not desirable, that an individual processor have data from disjoint regions of the simulation.  In that case each processor would have multiple disjoint sections of the AMR tree, but these disjoint sections would collectively be considered the processor's sub-tree.

Each time new grids on level $l+1$ are created (by local parent grids on level $l$), each processor determines how the new child grids should be distributed (i.e. which processor should get the new grids).  This distribution is carried out in the manner described in section \ref{loadbalancing} below.  Connections between the new level $l+1$ nodes and the rest of the tree must then be formed.  Because of the inheritability of the neighbor/overlap connections, even if a child grid is distributed to another processor, the connections between that child node and its neighbors/overlaps/parent are first made on the processor that created the grid (ie the processor containing the grid's parent).  If a processor's local grid has a remote parent, then that processor will always receive information about that local grid's neighbors/overlaps from the processor containing the remote parent.  This is true of neighbor and preceding overlap connections of new grids as well as succeeding overlap connections of old grids.

Before processors containing parents of level $l+1$ grids (both new and old) can send connection information to remote children (if they exist), these processors must first share information about the creation of children with each other.  Neighbor connections between new nodes on level $l+1$ require each processor to cycle through its local level $l$ nodes and identify those with remote neighbors living on other processors.  Once remote neighbors have been identified the information about new children from the local nodes is sent to the processor(s) containing the remote neighbors.  Not all children need to be sent to every remote neighbor.  Only those that are close enough to potentially be adjacent to the remote neighbor's children are necessary.  The information must flow in both directions meaning a individual processor also needs to {\it receive} imformation about potential new children from all other processors containing remote neighbors.

\section{Threaded Multilevel Advance} \label{threading}

\begin{figure}
 \caption{Plot showing threads of AMR algorithm}
 \centering
  \includegraphics[width=.84\textwidth]{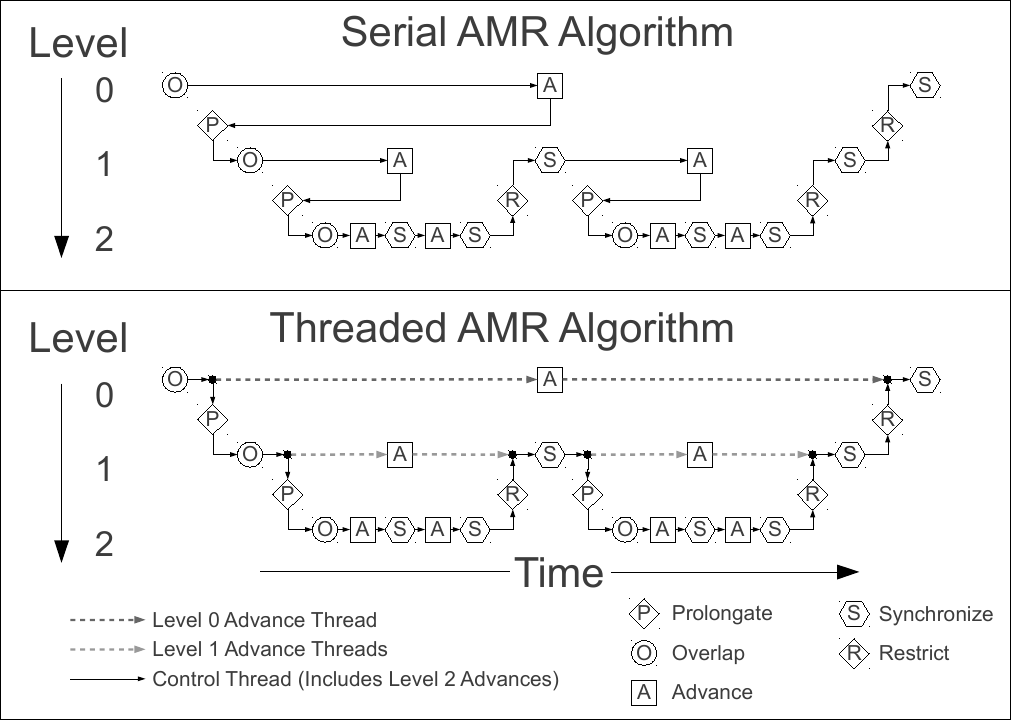} 
\label{threadingfig}
\end{figure}

 Many if not all current AMR codes tend to perform grid updates across all levels in a prescribed order that traverses the levels of the AMR hierarchy in a sequential manner.  Thus the codes begins at the base grid (level 0), moves down to the highest refinement level and then cycles  up and and down across levels based on time step and sychronization requirements(for a simulation with 3 levels the sequence would be: 0, 1, 2, 2, 1, 2, 2, 0...) In the top panel of figure \ref{threadingfig} the basic operations of (P)rolongating, (O)verlapping, (A)dvancing, (S)ynchronizing, and (R)estricting are shown for each level along with the single (sereal) control thread.  
 
 Good parallel performance requires each level update to be independently balanced across all processors (or at least levels with a significant fraction of the workload).  Load balancing each level, however, requires the levels to contain enough grids to be effectively distributed among the processors.  Such a requirement demands each level to be fairly "large" in the sense of having many grids or allowing each level's spatial coverage be artificially fragmented into small pieces.  The former situation leads to broad simulations (large base grid leaving resources for only a few levels of AMR), while the later situation leads to inefficient simulations due to the fair amount of overhead required for ghost zone calculations.

In the bottom panel of figure \ref{threadingfig} we show a schematic of the AstroBEAR 2.0 AMR algorithm.  In this figure basic operations of (P)rolongating, (O)ver-lapping, (A)dvancing, (S)ynchronizing, and (R)estricting are shown again.  This time however the level advances are independent and exist on separate threads of computation computation.  There is an overarching control thread which handles all of the communications and computations required for prolongation, overlapping, synchronizing, and restricting as well as the finest level advances.  Each coarser level advance has its own thread and can be carried forward independently with preference being given to the threads that must finish first (which is always the finer level threads).  In addition to relaxing the requirement of balancing every level, the existence of multiple threads allows processors to remain busy when the control thread becomes held up because it needs information from another processor.  For example, while waiting for ghost zone data for level 3 it can work on advancing levels 2, 1, or 0.

\section{Load Balancing} \label{loadbalancing}
As was discussed above, threading level advances removes the need for balancing each level independently and instead allows for global load balancing.  It also and perhaps more importantly allows for consideration of the progress of coarser advance threads when successively distributing the work load of finer grids.  This ``dynamic load balancing'' allows adjustments to be made to finer level distributions to compensate for variations in progress made on ongoing coarser advances.  When distributing grids on level $l$, the distribution is adjusted so that the total predicted remaining work load on level $l$ and coarser, is constant across processors.  More formally if $g_l^p$ is the current workload on processor $p$ and level $l$ and if $w_l^p$ is the current part of $g_l^p$ that has already been completed and $s_l$ is the number of remaining level $l$ steps in the entire AMR step, then we can compute the predicted remaining work load on all levels 0 through $l$ as $\eta_l^p=\displaystyle \sum_{l'=0}^{l}{s_{l'}g_{l'}^p-w_{l'}^p}$.  The successive distributions of $g_l^p$ are given by 
$ g_l^p=\overline{g_l}-\frac{\eta_{l-1}^p-\overline{\eta_{l-1}}}{s_l}$ so that $\eta_{l}^p = \overline{\eta_l}$.

 \section{Performance Results} \label{performanceresults}
 
  For our weak scaling tests we advected a magnetized cylinder across the domain until it was displaced by 1 cylinder radius.  The size of the cylinder was chosen to give a filling fraction of approximately 12.5\% so that in the AMR run, the work load for the first refined level was comparable to the base level.  The resolution of the base grid was adjusted to maintain $64^3$ cells per processor and we found that our ``out of the box`` weak scaling for both fixed grid and for AMR was better then 80\% out to 2048 processors.  We have not yet tested the performance on larger clusters.

\bibliography{author}{}

\end{document}